\documentclass[prd,aps,twocolumn,nopacs,superscriptaddress,nofootinbib]{revtex4}
\usepackage{graphicx}
\usepackage{dcolumn}
\usepackage{bm}
\usepackage{amsmath}
\usepackage{amssymb}
\usepackage{epsfig}
\usepackage{color}

\def\ii{{\rm i}}  \def\ee{{\rm e}}
\def\rb{{\bf r}}      
\def\xx{\hat{\bf x}}  \def\yy{\hat{\bf y}}  \def\zz{\hat{\bf z}}  
\def\kb{{\bf k}}    
    
\def\me{m_{\rm e}}  \def\kB{{k_{\rm B}}}
  \def\Bb{{\bf B}}  
\def\pb{{\bf p}}  
  
\def\uu{\hat{\bf u}}    \def\eeh{\hat{\bf e}}
\def\fe{\gamma^{\text{e}}}  \def\fa{\gamma^{\text{a}}}
\def\HI{{\mathcal{H}_{\rm I}}}    \def\wc{\omega_{\rm c}}

\begin{document}

\title{Magnetically Activated Rotational Vacuum Friction}
\author{Deng~Pan}
\affiliation{ICFO-Institut de Ciencies Fotoniques, The Barcelona Institute of Science and Technology, 08860 Castelldefels (Barcelona), Spain}
\author{Hongxing~Xu}
\affiliation{School of Physics and Technology, Wuhan University, Wuhan 430072, China}
\author{F.~Javier~Garc\'{\i}a~de~Abajo}
\email[Corresponding author: ]{javier.garciadeabajo@nanophotonics.es}
\affiliation{ICFO-Institut de Ciencies Fotoniques, The Barcelona Institute of Science and Technology, 08860 Castelldefels (Barcelona), Spain}
\affiliation{ICREA-Instituci\'o Catalana de Recerca i Estudis Avan\c{c}ats, Passeig Llu\'{\i}s Companys 23, 08010 Barcelona, Spain}

\begin{abstract}
We predict the existence of a torque acting on an isotropic neutral nanosphere activated by a static magnetic field when the particle temperature differs from the surrounding vacuum. This phenomenon originates in time-reversal symmetry breaking of the particle interaction with the vacuum electromagnetic field. We present a rigorous quantum treatment of photons and particle excitations that leads to a nonzero torque even in a motionless particle. We also find that the dynamical evolution of the particle temperature and rotation frequency follow an exotic dynamics, including spontaneous changes in the rotation direction. Magnetically activated thermal vacuum torques open a unique avenue for the investigation of the effect of time-reversal symmetry-breaking in thermal and Casimir physics.
\end{abstract}
\date{\today}
\maketitle

\section{Introduction}

Coupling between the bosonic excitations of moving objects (e.g., plasmons or phonons) and the vacuum electromagnetic field can produce net transfers of momentum and emission of real photons at the expense of mechanical motion \cite{M1970_3,CF1984,C92,L94,SSW94,LJR96,SH96,P97,
P98_2,KG99,L99,BMM01,P05,L05_2,P06,YO08,D10,P10,
B10_3,B10_4,paper157,paper166,DNM11,M11_2,
VP11,WJP11,paper172,DD12,LDJ12,MJK12,paper199,LPH13,
MGK13,MGK13_2,WLL14,BL15,JLD15,IBH16,MRC17,
STB17,paper289}.
These phenomena have been explored in accelerated mirrors \cite{M1970_3,L94,SH96,YO08,WJP11,LPH13}, sliding surfaces \cite{P97,P98_2,P10,MGK13,WLL14,VP11}, rotating objects \cite{P05,P06,paper157,paper166,paper172,paper199,MJK12,paper289}, optical cavities \cite{LJR96,L05_2,DD12}, and moving atoms and particles \cite{B10_3,B10_4,LDJ12,BL15,MRC17,STB17}. For example, two planar homogeneous surfaces in relative parallel motion undergo contactless friction due to exchanges of surface excitations that interact through the vacuum electromagnetic field \cite{P97,P10}. Friction can additionally occur by emitting photon pairs if the two media are transparent and their relative velocity exceeds the Cherenkov condition \cite{P98_2,MGK13,S14}. The continous change in the dielectric boundaries associated with the rotation of a nonspherical object made of a nonabsorbing material also leads to stopping, assisted by the emission of photon pairs \cite{P05,P06}. More intriguing is the case of a spinning lossy sphere: despite the apparent preservation of dielectric boundaries, it undergoes a frictional torque even when the entire system is at zero temperature \cite{ZRS1986,paper157}, while the torque can be enlarged by the presence of a planar surface \cite{paper199}, giving rise to a lateral force \cite{paper289}.

Vacuum friction is closely related to time-reversal symmetry (\emph{T}-symmetry) of the electromagnetic field in the vicinity of the involved materials. Considering again two moving parallel surfaces \cite{P10}, \emph{T}-symmetry implies that the excitations of one of them have equal local density of states (LDOS) in its rest frame regardless of the orientations of their wave vectors. However, \emph{T}-symmetry is broken for the surrounding electromagnetic field due to the Fresnel drag associated with the moving surface, a result that has been recently exploited to design optomechanically-induced nonreciprocal optical devices \cite{MRL09,HR12,WZG13,RMA16,SZC16}. \emph{T}-symmetry breaking is a direct consequence of the different Doppler shifts experienced by excitations propagating along opposite directions in the moving surface, which understandably exhibit a LDOS asymmetry. This produces an imbalance in the momentum exchanged during transfers of excitations between the two surfaces, giving rise to a net stopping force. In a similar fashion, the rotational Doppler effect in a rotating spherical particle induces \emph{T}-symmetry breaking between excitations circulating in clockwise and anticlockwise directions, which also results in a vacuum frictional torque. From these general considerations, one would expect the emergence of vacuum forces in geometrically symmetric structures composed of nonreciprocal materials, in which \emph{T}-symmetry is broken for example by applying a static magnetic field.

\begin{figure}
\begin{centering}
\includegraphics[width=0.5\textwidth]{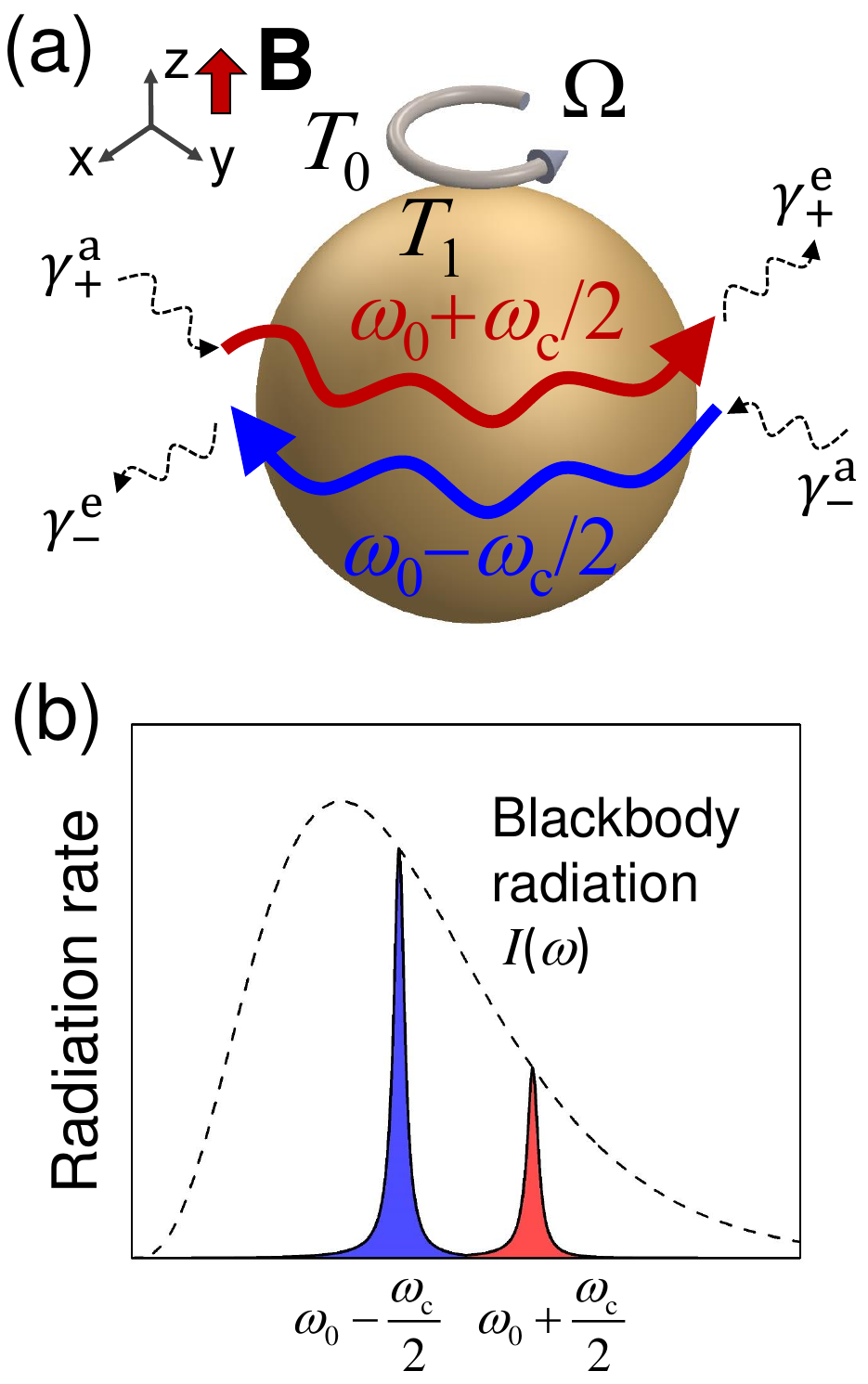}
\par\end{centering}
\caption{Imbalance of photon absorption and emission caused by Zeeman splitting in a rotating nanoparticle. ({\bf a}) We consider a sphere rotating with frequency $\Omega$ around the direction of a static magnetic field $\Bb\parallel\zz$. The temperatures of the particle and the surrounding vacuum are $T_1$ and $T_0$, respectively. A dipolar excited state in the sphere undergoes a splitting $\wc=eB/m_{\rm e}c$. Emission and absorption rates $\gamma_{\pm}^{\rm e}$ and $\gamma_{\pm}^{\rm a}$ from these two states are also affected by Doppler shift  associated with rotation. ({\bf b}) When 􏰉$\Omega=0$ and $T_0=0$, the asymmetric coupling of the excited states to radiation is determined by the blackbody distribution $\propto\omega^3n_1(\omega)$ at temperature $T_1$ (dashed curve).}
\label{Fig1}
\end{figure}

In the present paper we show that a spherical particle experiences a counterintuitive torque due to \emph{T}-symmetry breaking induced by a static magnetic field. We formulate a rigorous quantum-electrodynamic model to describe the system and show that a finite torque is exerted parallel to the magnetic field even on a motionless particle, provided its temperature differs from that of the surrounding vacuum. The torque originates in the asymmetric thermal population of particle internal bosonic excitation modes with opposite angular momentum (AM). We find the particle temperature and rotation frequency to follow an exotic dynamics characterized by spontaneous changes in the direction of rotation. We anticipate that similar vacuum forces should generally appear in nonmagnetic nanostructures when optical \emph{T}-symmetry is broken by means of static magnetic fields.

\section{Calculation of the frictional torque and absorption power}
\label{sec2}

We present a self-contained derivation of the torque and the absorption power that is general for a particle with axial symmetry and a static magnetic field along the rotation axis, as shown by a sphere in Fig.\ \ref{Fig1}(a). For motionless particles ($\Omega=0$) with axial symmetry, the magnetic field induces a splitting equal to the cyclotron frequency $\wc$ on the resonance peaks of the two optical dipolar modes that are both initially located at $\omega_0$, as illustrated in Fig.\ \ref{Fig1}(a). In fact, when focusing on particles that possess isotropic dipolar polarizability in the plane perpendicular to the magnetic field, the magnitude of this magnetic splitting depends on the actual morphology. The magnetic splitting of $\wc$ in an axially-symmetric particle originates in the Zeeman shifts of its internal electron states, which are quantized in accordance with the matrix elements of the orbital momentum operator \cite{M1966}. In contrast, in a cross-like particle, these matrix elements can be neglected in the internal electronic states and the magnetic field does not affect its optical response. In this study, we focus on particles with axial symmetry.

Our derivation also shows that a rotating particle experiences a gyromagnetic effect \cite{B1915,B1935_2,HJ1962} in the frame rotating with it, where the Coriolis forces acting on the particle electrons can be assimilated to an effective magnetic field. In particular, when such electronic states are eigenstates of the AM operator (e.g., in ellipsoidal particles), the excitation
modes in the polarizability display an additional frequency
splitting equal to $2\Omega$ in the rotating frame, but do not
undergo any correction due to rotation in the lab frame (i.e., it is the same regardless of the rotation speed). This is in contrast to previous results \cite{paper157,paper166,paper199,MJK12,MGK13,MGK13_2,MJK14,LS16,paper289}, in which the resonance peaks in the particle polarizability was frequency shifted in the lab frame due to rotation, a possibility that only holds for particles whose polarizabilities are not affected by this effective magnetic field (e.g., in cross-like particles, for which a classical description based on rigid rods also generates this correction \cite{paper172}, essentially due to the inability of Coriolis forces to produce motion transversal to the rods).

Incidentally, at low optical frequencies $\omega$, absorption in metallic particles is dominated by Ohmic losses and scales linearly with $\omega$, but for this term the AM correction becomes negligible and the imaginary part of the polarizability needs to be corrected due to rotation, leading to a term proportional to $\omega-\Omega$ regardless of particle morphology (see results for this limit in Refs.\ \cite{paper157}). However, we are concerned here with the effect of optical resonances in rotating particles under the influence of a magnetic field, for which Ohmic losses can be safely ignored.

\subsection{Quantum states in a rotating particle}
\label{IIA}

We study an axially-symmetric nanoparticle rotating with angular frequency $\Omega$ at temperature $T_1$ in a vacuum at temperature $T_0$, exposed to a static magnetic field. The torque on this particle originates in the energy and AM exchanges between its excitations and the vacuum electromagnetic field, as illustrated in Fig.\ \ref{Fig1}(a). We assume the rotation direction and the magnetic field to be aligned along the axis of symmetry of the particle (direction $z$). The internal electronic states of the particle $\left|l\right\rangle$ of energies $\hbar\varepsilon_l$ can be chosen to be associated with azimuthal numbers $m_l$ corresponding to a rotational wave function $\ee^{\ii m_l\varphi}$, where $\varphi$ is the azimuthal angle around $z$. We remark that the present formalism applies to any type of electronic states, including many body descriptions of the problem.

In the lab frame and in the absence of a magnetic field, the energies of the quantum states $\left|l\right\rangle$ are $\hbar\varepsilon_l^0$. The applied magnetic field $B$ induces Zeeman shifts $m_l\wc/2$, where $\wc=eB/m_{\rm e}c$ is the cyclotron frequency, so the state frequencies become $\varepsilon_l=\varepsilon_l^0+m_l\wc/2$.

For a rotating particle, it is important to consider the states in the frame rotating with the particle because thermal equilibrium must be established among those states. Under a transformation from the lab frame to the rotating frame, $\rb=(r,\varphi,z,t)\rightarrow\rb'=(r',\varphi',z',t')=(r,\varphi-\Omega t,z,t)$, taking into account that the time derivative transforms as $\partial_t\rightarrow\partial_{t'}-\Omega\partial_{\varphi'}$, the Schr\"odinger equation in the rotating frame becomes ($\mathcal{H}_{\rm part}-\Omega\mathcal{L}_z)\psi=\ii\hbar\partial_{t'}\psi$ \cite{RR04}, where $\mathcal{H}_{\rm part}$ is the particle Hamiltonian in the lab frame (with $\rb$ substituted by $\rb'$), $\mathcal{L}_z=-\ii\hbar\partial_{\varphi'}$ is the AM operator around the rotation axis, and the term $-\Omega\mathcal{L}_z$ introduces an additional shift in the frequencies of the quantum states, which now become $\varepsilon_l-m_l\Omega$ (see Appendix\ \ref{A1}). This term can be regarded as the result of an effective magnetic field acting on the particle in the rotating frame. When thermal equilibrium is established among the states in the rotating frame, the populations of quantum states $\left|l\right\rangle$ are then determined by their energies $\hbar(\varepsilon_l-m_l\Omega)$ according to the Boltzmann distribution $f_l=\ee^{-\hbar(\varepsilon_l-m_l\Omega)/\kB T_1}/Z_1$, where $Z_1$ is the partition function in the rotating frame.

We note that in previous works \cite{paper157,paper166,paper199,MJK12,MGK13,MGK13_2,MJK14,LS16,paper289} on rotational vacuum friction, the correction $-\Omega\mathcal{L}_z$ is missing, so they only apply to particles in which the matrix elements of $\mathcal{L}_z$ are negligible.

We remark that for particles that possess isotropic dipolar polarizability in the plane perpendicular to the rotation axis, these matrix elements depend on the actual morphology. For example, the matrix elements are small in cross-like structures. For particles with axial symmetry, the effect of this term needs to be incorporated, as explained in the present paper. In particular, when such electronic states are eigenstates of the AM operator (e.g., in ellipsoidal particles), the excitation modes in the polarizability do not undergo any correction due to rotation in the lab frame (i.e., it is the same regardless
of rotation speed).

\subsection{Particle-vacuum interaction}
\label{interaction}

We consider a complete basis set of vacuum photon modes $i$ labeled by the occupation numbers $n_i$. The state of the particle-field system is thus expressed as a combination of states $|l,\{n_i\}\rangle$, where $l$ labels the internal electronic states of the particle (see Sec.\ \ref{IIA}). Photons and electronic excitations are coupled through the interaction Hamiltonian \cite{paper166}
\begin{align}
\HI=-\sum_{i} \sqrt{\frac{2\pi\hbar\omega_i}{V}}\;\eeh_i\left(a_i^++a_i\right)\cdot\pb,
\label{HI}
\end{align}
where $V$ is the quantization volume, $a_i$ and $a_i^+$ are the annihilation and creation operators of a photon in mode $i$, respectively, $\omega_i$ and $\hat{\bf e}_i$ are the frequency and (real) unit polarization vector of the photon, and $\pb$ is the particle dipole operator. By describing particle-photon interactions through the excitation dipoles, we are assuming that the particle is small compared with the wavelengths of the involved photons.

When considering transitions in the particle driven by $\HI$, we need to evaluate dipole matrix elements $\pb_{l'l}=-e\langle l'|\rb|l\rangle$, for which it is convenient to use the Zeeman coordinate basis, defined by the unit vectors $\uu_\pm=(\xx\pm\ii\yy)/\sqrt{2}$. Taking into account that the $|l\rangle$ states have well-defined AM $\hbar m_l$, the Zeeman basis readily reveals the selection rule $m_{l'}=m_l\pm1$ for polarization in the $x$-$y$ plane and  $m_{l'}=m_l$ for polarization along $z$. More precisely, we can write $\pb_{l'l}=p^\perp_{l'l}\,\left(\uu_+\delta_{m_{l'},m_l-1}+\uu_-\delta_{m_{l'},m_l+1}\right)+p^z_{l'l}\,\zz\delta_{m_{l'},m_l}$, where $p^z_{l'l}=\zz\cdot\pb_{l'l}$ and $p^\perp_{l'l}=\uu_\pm\cdot\pb_{l'l}$ is independent of the sign of $m_{l'}-m_l$ due to rotational degeneracy. Using these expressions, the only nonzero matrix elements of $\HI$ are
\begin{subequations}
\begin{align}
\langle l',n_i+1|\HI|l,n_i\rangle &= -\sqrt{\frac{2\pi\hbar \omega_i}{V}}\;\Delta_{l'l,i}\sqrt{(n_i+1)}, \label{Deltalla}\\
\langle l',n_i-1|\HI|l,n_i\rangle &= -\sqrt{\frac{2\pi\hbar \omega_i}{V}}\;\Delta_{l'l,i}\sqrt{n_i}, \label{Deltallb}
\end{align}
\label{Deltall}
\end{subequations}
where
\begin{align}
\Delta_{l'l,i}=
p^\perp_{l'l}\left(e_i^+\delta_{m_{l'},m_l+1}+e_i^-\delta_{m_{l'},m_l-1}\right)+p^z_{l'l}e_i^z\; \delta_{m',m} \nonumber
\end{align}
and $e_i^\pm=\uu_\mp\cdot\eeh_i$.

Because the dipole components along $z$ cannot produce changes in AM, we only need to consider polarization in the $x$-$y$ plane in the calculation of the torque. However, we need to account for polarization along $z$ when calculating the particle absorption power (see below). Using the Fermi golden rule with the above transition matrix elements, we can now obtain the rates $\gamma_\pm^{\rm e}$ (contribution arising from $n_i\rightarrow n_i+1$ and $m_{l'}-m_l=\mp1$ terms) and $\gamma_\pm^{\rm a}$ (from $n_i\rightarrow n_i-1$ and $m_{l'}-m_l=\pm1$) associated with photon emission and absorption processes, and accompanied by a net particle AM change given by $\mp\hbar$ and $\pm\hbar$, respectively; these rates are thus separated in components with opposite AM as shown in Fig.\ \ref{Fig1}(a). In particular, the absorption rates reduce to
\begin{align}
\gamma_{\pm}^{\rm a}&=\frac{2\pi}{\hbar^2} \sum_{ll'} f_l \sum_i\sum_{n_i}^{\infty} \frac{\ee^{-n_i\hbar\omega_i/\kB T_0}}{Z_{0,i}} \; \delta_{m_{l'},m_l\pm1}
\nonumber\\
&\;\;\;\;\;\;\;\;\;\;
\times \left|\left\langle \l',n_i-1|\HI|l,n_i\right\rangle\right|^2
\delta(\varepsilon_{l'l}-\omega_i),\nonumber
\end{align}
where $\varepsilon_{l'l}=\varepsilon_{l'}-\varepsilon_l$ and we perform the thermal average over electronic states $\left|l\right\rangle$ (initial populations $f_l$) and photon states $\left|n_i\right\rangle$. Here, $Z_{0,i}=\sum_{n_i}\ee^{-n_i\hbar\omega_i/\kB T_0}$ is the partition function of photon mode $i$, assumed to be at thermal equilibrium for a vacuum temperature $T_0$. Now, noticing that the $\HI$ matrix elements of Eqs.\ (\ref{Deltall}) introduce terms proportional to $n_i$ and $n_i+1$ in the rates, the sum over $n_i$ readily reduces to factors proportional to $n_0(\omega_i)$ and $n_0(\omega_i)+1$, respectively, where $n_0(\omega)=1/(\ee^{\hbar\omega/\kB T_0}-1)$ is the Bose-Einstein distribution function at the vacuum temperature. Also, we perform the sum over $i$ by using a plane-wave representation of the photon states and making the substitution $\sum_i \rightarrow V/(2\pi)^3\sum_\sigma\int d^3\kb$, where $\sigma$ and $\kb$ denote photon polarization and wave vector. The angular part of this integral can be conveniently carried out using azimuthal and polar vectors as the two orthogonal polarization states for each direction of $\kb$. We are then left with an integral over the photon frequency, which reduces the absorption rates to
\begin{align}
\gamma_{\pm}^{\rm a}=\frac{4\pi^2}{\hbar}
\sum_{ll'} &f_l (p^\perp_{l'l})^2\,\delta_{m_{l'},m_l\pm 1}
\nonumber\\
&\times\int_0^\infty  \omega\rho^0(\omega)d\omega\, n_0(\omega)\delta(\varepsilon_{l'l}-\omega),\nonumber
\end{align}
where $\rho^0(\omega)=\omega^2/3\pi^2 c^3$ is the projected local density of optical states in vacuum \cite{paper102}. Finally, we compare this expression to the polarizability of a rotating particle, which is calculated in Appendix\ \ref{secpola} using the notation and states introduced in Sec.\ \ref{IIA} [see Eq.\ (\ref{Imalpha})]. The result is
\begin{align}
\gamma_{\pm}^{\rm a}=4\pi \int_0^\infty \omega \rho^0(\omega) d\omega\, & n_0(\omega)[n_1(\omega\mp\Omega)+1]
\label{AB} \\
&\times{\rm Im} \{\alpha_\pm(\omega)\}, \nonumber
\end{align}
where
\begin{align}
\alpha_\pm(\omega)&=\frac{1}{\hbar}\sum_{ll'} (f_{l'}-f_l) \delta_{m_{l'},m_l\pm1}\frac{(p^\perp_{l'l})^2}{\omega-\varepsilon_{l'l}+\ii 0^+} \label{alphapm}
\end{align}
is the polarizability for circularly polarization in the $x$-$y$ plane. Proceeding in a similar way, using Eq.\ (\ref{Deltallb}) instead of Eq.\ (\ref{Deltalla}), we find the emission rates associated with a net AM change $\pm\hbar$ in the particle as
\begin{align}
\gamma_{\pm}^{\rm e}=4\pi \int_0^\infty  \omega \rho^0(\omega)d\omega\, & [n_0(\omega)+1]n_1(\omega\mp\Omega)
\label{EM} \\
&\times{\rm Im} \{\alpha_\pm(\omega)\}. \nonumber
\end{align}
We observe that emission and absorption processes, as  described in Eqs. (\ref{AB}) and (\ref{EM}), incorporate a Bose-Einstein statistics for both photons and particle excitations, introduced through their distributions $n_0(\omega)$ and $n_1(\omega)$ at temperatures $T_0$ and $T_1$, respectively, which enter as factors $n_j+1$ or $n_j$ depending on whether an excitation is added to or removed from the particle ($j=1$) or the vacuum ($j=0$). This is remarkable considering that we have not made any assumptions regarding the statistics of the particle excitations, so it can equally apply to a discrete set of many-body states $|l\rangle$ or to bosonic modes such as plasmons.

\subsection{Torque and absorption power}

We are now ready to calculate the torque acting on the particle by summing the above transition rates, multiplied by their respective transferred AM $(m_{l'}-m_l)\hbar$. We find
\begin{align}
M&=\hbar[(\fe_- -\fa_-) -(\fe_+ -\fa_+)]\nonumber\\
&=4\pi\hbar\sum_{\nu=\pm1} \nu\int_0^\infty\omega \rho^0(\omega) d\omega\,
N_\nu(\omega) \;{\rm Im}\{\alpha_\nu(\omega)\}  \label{M}
\end{align}
where
\[N_\nu(\omega)=n_0(\omega)-n_1(\omega-\nu\Omega)\]
is the imbalance of vacuum and particle mode populations. We point out the presence of ${\rm Im}\{\alpha_\nu(\omega)\}$ in Eq.\ (\ref{M}), in contrast to the expression ${\rm Im}\{\alpha_\nu(\omega-\Omega)\}$ used in previous works \cite{paper157,paper166,paper199,MJK12,MGK13,MGK13_2,MJK14,LS16,paper289}; the contribution of $\mathcal{L}_z$ to the energy differences in the electronic transitions that configure the dipolar polarizability cancel exactly the rotational Doppler shift in $\Omega$ when the electronic states are eigenstates of $\mathcal{L}_z$ (see Appendix\ \ref{secpola}), which is the case for the particles with rotational symmetry here discussed. The interactions between the vacuum field and the two particle modes carrying a difference of AM given by $\pm\hbar$ follow the general principle that the radiative exchange with photon (i.e., boson) modes is proportional to the imbalance of their thermal populations. However, for the rotating particle under consideration, the populations of the particle excitations are evaluated at Doppler-shifted frequencies in the rotating frame (i.e., $\omega\mp\Omega$). As discussed in Ref.\ \cite{paperPan}, a solid particle rotating at low frequency $\Omega$ in the absence of a magnetic field does not show optical dichroism [i.e., $\alpha_+(\omega)=\alpha_-(\omega)$], and consequently, the rotational friction torque only arises from a population imbalance driven by the Doppler shift, which causes $n_1(\omega+\Omega)-n_1(\omega-\Omega)$ to be nonzero, unless $\Omega=0$.

For a motionless particle ($\Omega=0$), a nonzero torque can also emerge as the effect of circular dichroism of the particle in the presence of a DC magnetic field. For example, with the vacuum at zero temperature ($T_0=0$), photon emission is produced at rates proportional to the blackbody spectrum $\propto \omega^3 n_1(\omega)$ [Fig.\ \ref{Fig1}(b)], while absorption is obviously zero; additionally, the rotational invariance of the particle implies that it possesses at least two excitation modes with opposite AM [i.e., $\pm\hbar$, see red and blue arrows in Fig.\ \ref{Fig1}(a)], therefore undergoing opposite Zeeman shifts, which in turn results in an imbalance of ${\rm Im}\{\alpha_\nu(\omega)\}$ at the split frequencies of the two modes; as a result of this, the contribution to the torque associated with emission from each of the two modes does not cancel completely [see Fig.\ \ref{Fig1}(b)], leaving a net contribution. We thus predict that a motionless particle should experience a nonzero torque in the presence of a DC field when its temperature differs from the vacuum.

We can also use the above rates to obtain the power absorbed by the particle, considering that each photon absorption or emission involves a particle energy gain or loss given by $\hbar \omega$. The absorption power then reduces to
\begin{align}
P^{\rm abs}=4\pi\hbar\sum_{\nu=0,\pm} \int_0^\infty\omega^2 \rho^0(\omega) d\omega\,
 N_\nu(\omega) \;{\rm Im}\{\alpha_\nu(\omega)\},  \label{P}
\end{align}
where we have included a term $\nu=0$ associated with power exchanges due to particle polarization along the rotation direction $z$ (see Appendix\ \ref{secpola}).

\section{Torque and dynamics in the presence of a magnetic field}

The presence of a magnetic field produces frequency shifts $m_l\wc/2$ in the particle excited states $|l\rangle$, depending on their AM number $m_l$. This directly affects the polarizability $\alpha_\pm$ [Eq.\ (\ref{alphapm})] through the frequency differences $\omega_{l'l}=\pm\wc/2$, leading to a non-reciprocal response characterized by $\alpha_+\neq\alpha_-$. As a direct consequence of this, Eq.\ (\ref{M}) yields a nonzero torque even in the absence of rotation ($\Omega=0$), provided $T_1\neq T_0$.

\begin{figure}
\begin{centering}
\includegraphics[width=0.5\textwidth]{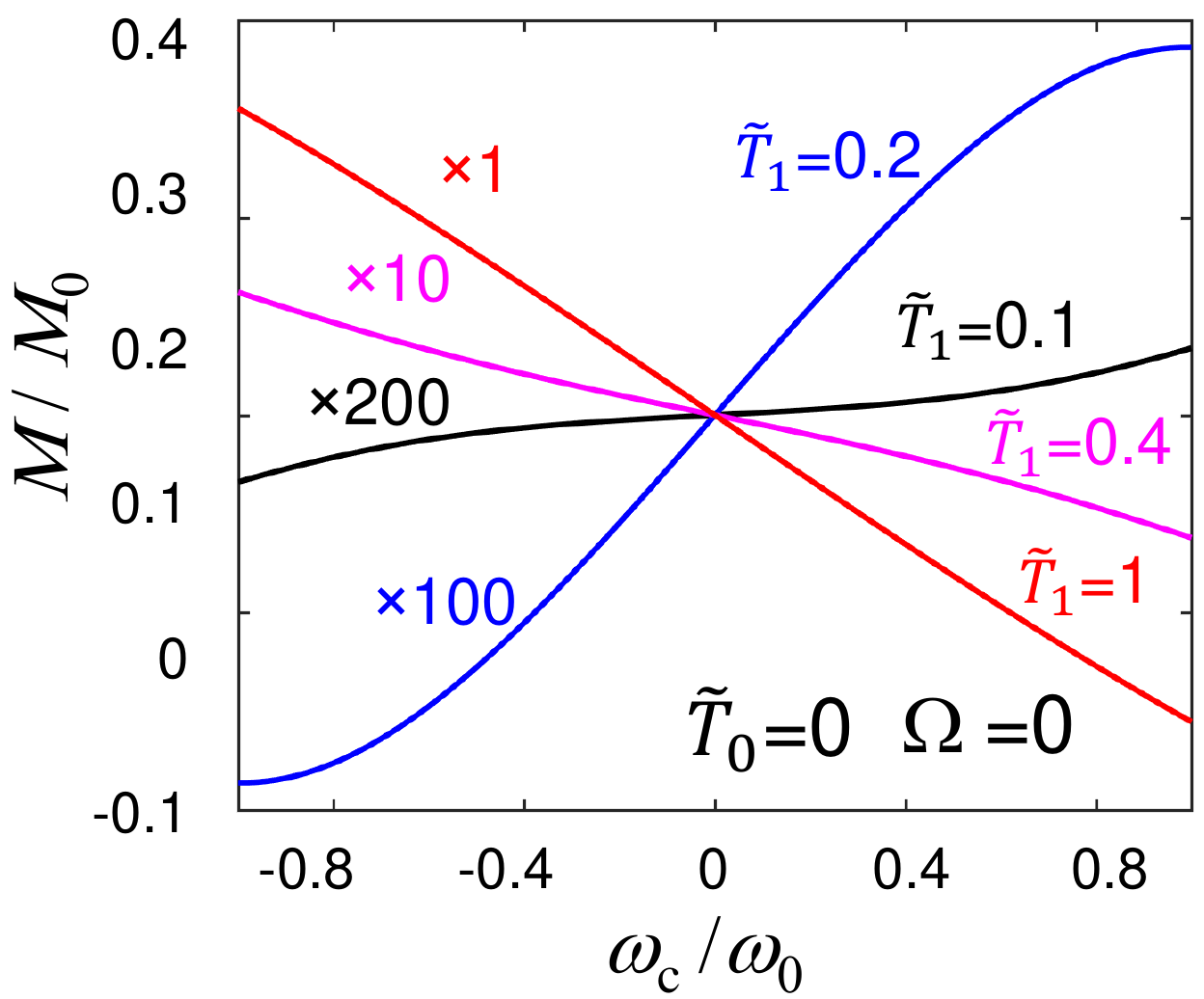}
\par\end{centering}
\caption{Torque experienced by a motionless nanosphere ($\Omega=0$) in the presence of a magnetic field. We present the torque as a function of magnetic field $B$, quantified through the cyclotron frequency $\wc=eB/m_{\rm e}c$, for a particle characterized by a dipolar excitation mode at frequency $\omega_0$. Results are shown for different normalized particle temperatures $\tilde{T}_1=\kB T_1/\hbar\omega_0$. The torque is given in units of $M_0=\hbar\gamma_0$, where $\gamma_0$ is the natural radiative decay rate of the $\omega_0$ particle mode. The vacuum is at temperature $T_0=0$.}
\label{Fig2}
\end{figure}

\begin{figure}
\begin{centering}
\includegraphics[width=0.5\textwidth]{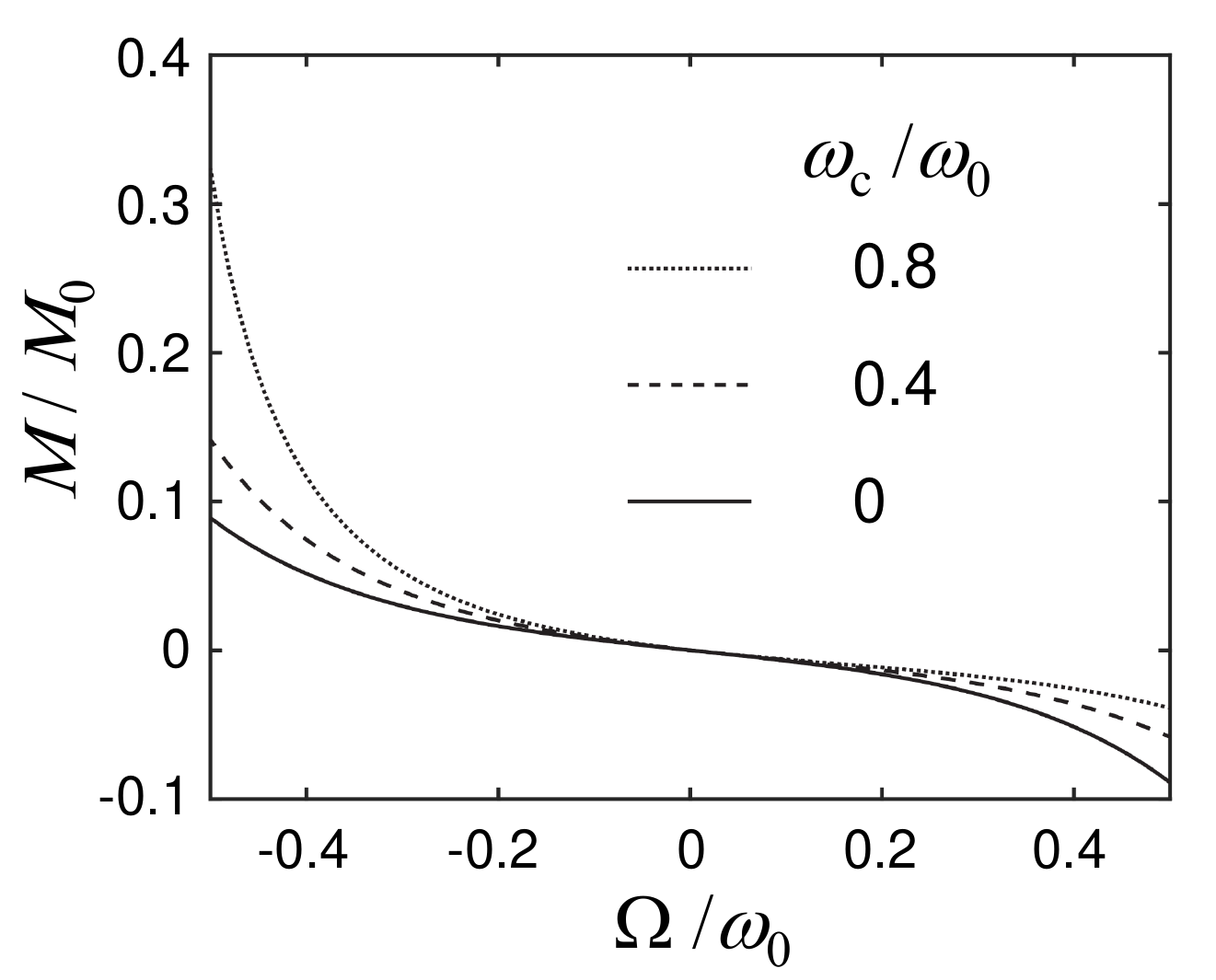}
\par\end{centering}
\caption{Frictional torque acting on a rotating nanosphere under a static magnetic field as a function of rotation frequency for different magnetic splittings $\wc$. We assume the particle and the vacuum to be at the same temperature equal to $T_j=0.2\,\hbar\omega_0/\kB$. All other parameters are the same as in Fig.\ \ref{Fig2}.}
\label{Fig3}
\end{figure}

\begin{figure}
\begin{centering}
\includegraphics[width=0.5\textwidth]{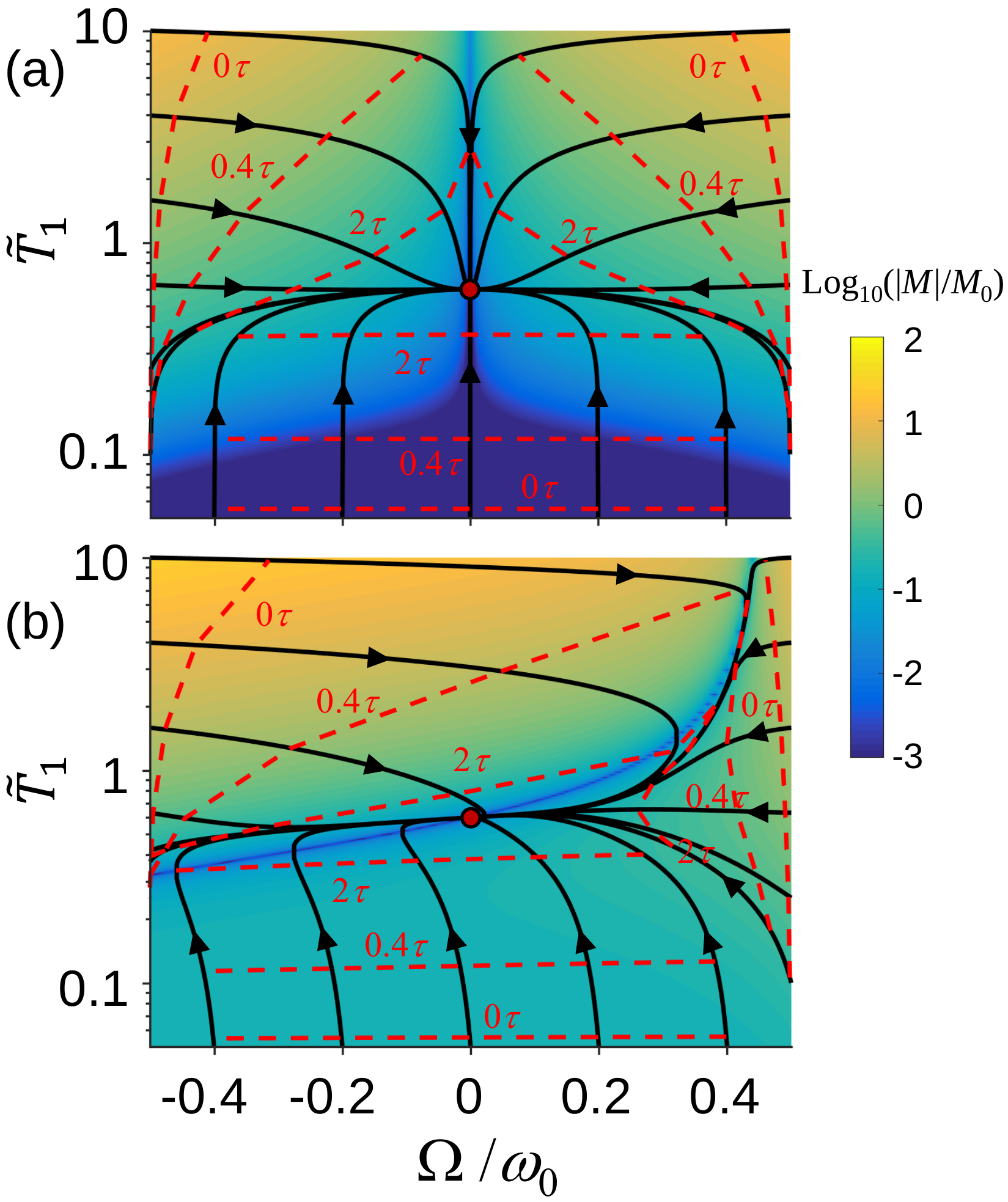}
\par\end{centering}
\caption{Temporal dynamics in the space of rotation frequency $\Omega$ and particle temperature $T_1$ for $\wc=0$ ({\bf a}) and $\wc=0.8\,\omega_0$ ({\bf b}). The normalized vacuum temperature is $\tilde{T}_0=0.6$. The dashed curves and corresponding numerical labels indicate evolution times in units of $\tau=I\omega_0/\hbar\gamma_0$ for $C/I=\kB\omega_0/\hbar$, where $I$ is the moment of inertia and $C$ is the heat capacity of the particle. A red circle indicates the equilibrium point $M=0$ and $P=0$.}
\label{Fig4}
\end{figure}

Although the results presented in Sec.\ \ref{sec2} are general, in what follows we focus for simplicity on an isotropic sphere characterized by a degenerate dipolar mode of frequency $\omega_0$. Also, we find it convenient to use the natural radiative decay rate \cite{L1983} $\gamma_0=4\omega_0^3p_0^2/3\hbar c^3$ as a parameter, instead of the excitation dipole moment of the mode $p_0$. We further assume the mode width to be small compared with $\omega_0$. Under these conditions, the particle polarizability in Eq.\ (\ref{M}) and (\ref{P}) reduces to ${\rm Im}\{\alpha_\pm(\omega)\}=(\pi p_0^2/\hbar)\delta(\omega-\omega_0\mp\wc/2)$ with $\pm$ components differing due to Zeeman splitting (see Appendix\ \ref{secpola}). Inserting these expression into Eq.\ (\ref{M}), we readily find
\begin{align}
M=\frac{\hbar\gamma_0}{\omega_0^3} \left[(\omega_0^+)^3 N_+(\omega_0^+)- (\omega_0^-)^3 N_-(\omega_0^-)\right]
\label{M2}
\end{align}
where $\omega_0^\pm=\omega_0\pm\wc/2$ are the resonance frequencies of the two Zeeman-split particle modes. Similarly the absorption power in Eq.\ (\ref{P}) reduces to
\begin{align}
P^{\rm abs}=\frac{\hbar\gamma_0}{\omega_0^3} \left[(\omega_0^+)^4 N_+(\omega_0^+)+ (\omega_0^-)^4 N_-(\omega_0^-)+\omega_0^4N_0(\omega)\right].
\label{P2}
\end{align}
We now use Eqs.\ (\ref{M2}) and (\ref{P2}) to produce the numerical results presented in Figs.\ \ref{Fig2}-\ref{Fig4}.

Figure\ \ref{Fig2} shows the torque on a motionless nanoparticle ($\Omega=0$) calculated from Eq.\ (\ref{M2}) for various nanoparticle temperatures $T_1$ as a function of magnetic field strength (cyclotron frequency $\wc$) when the vacuum is at temperature $T_0=0$. The rotational symmetry of the particle implies that the torque changes sign when the direction of the magnetic field $\Bb$ is reversed. The direction of the torque also depends on particle temperature: it is roughly parallel (anti-parallel) to $\Bb$ at low (high) $T_1$. This behavior is clearly illustrated by the expression $M=(\hbar\gamma_0/\omega_0^3)\left[(\omega_0^-)^3n_1(\omega_0^-)-(\omega_0^+)^3n_1(\omega_0^+)\right]$, which is valid for $T_0=0$, $\Omega=0$, and $|\wc/2|<\omega_0$; under the conditions of Fig.\ \ref{Fig1}(b) (low $T_1$), the high-energy mode at $\omega_0^+$ decays more slowly than the mode at $\omega_0^-$, and hence $M>0$, whereas the opposite behavior is observed when the state energies lie to the left of the emission maximum (high $T_1$).

We find it interesting that, despite the dipolar nature of the particle under consideration, the vacuum torque can be asymmetric with respect to sign changes in the rotation frequency (i.e., it not only changes sign, but also magnitude). An illustration of this effect is shown in Fig.\ \ref{Fig3}, which is obtained by using Eq.\ (\ref{M2}). In the absence of a magnetic field, the torque is symmetric, while the introduction of a magnetic field produces a sizable asymmetry. We attribute this effect to the optical nonreciprocity of the material under the influence of a magnetic field.

The frictional torque of a sphere in the presence of a magnetic field leads to exotic dynamics, as shown in Fig.\ \ref{Fig4}, where we study the evolution of particle rotation frequency $\Omega$ and temperature $T_1$ as a function of time. Because the ionic masses inside the sphere are large compared with the electron mass, the evolution of the particle dynamics is governed by the classical equations of motion $\dot{\Omega}=M/I$ for the rotation velocity and $\dot{T}_1=\left(P^{\rm abs}-M\Omega\right)/C$ for the temperature, where $I$ is the moment of inertia, $C$ is the heat capacity, and the term $M\Omega$ is the fraction of absorbed power that is converted into rotational energy rather than internal heating of the particle. A set of universal landscapes are found for the evolution of $\kB T_1/\hbar$ and $\Omega/\omega_0$ as a function of the fixed parameters $\wc/\omega_0$, $\kB T_0/\hbar$, and $C/I$. A particular numerical solution is plotted in Fig.\ \ref{Fig3}(b), where we observe evolution lines that are strongly influenced by the magnetic field. The system is shown to evolve toward the equilibrium point for all initial configurations. Interestingly, the evolution toward the equilibrium point is often involving stopping of the particle out of equilibrium and changes in the direction of rotation.

\section{Concluding remarks}

We conclude that the presence of a static magnetic field can lead to nontrivial torques acting on a nanoparticle when its temperature differs from that of the surrounding medium. We have derived the resulting torque by calculating the different rates of radiation exchange between particle excitations and the environment, including the thermal populations of excitations for opposite values of their AM, which differ due to Zeeman splitting produced by the magnetic field.

The emergence of thermal vacuum torques in nonmagnetic particles subject to static magnetic fields suggests a radically new way of mechanically controlling nanoscale objects. Remarkably, these torques exist even when the particle is nonrotating. The magnetic field also influences the dynamics of the system significantly. These findings could be explored by observing the dynamical evolution of small-particle gases (e.g., through rotational frequency shifts \cite{BB97,CRD98,MHS05}) held in vacuum inside a container that is subject to an external magnetic field. The sum of torques of an ensemble of particles contained inside a dielectric matrix could be also measured macroscopically. Additionally, one could use a low-frequency electric field polarized along the rotation axis to heat the particle and control its temperature, so that dynamical equilibrium is then established at a rotation frequency that depends on both the applied heating and the external magnetic field.  Larger torques $\propto\wc$ could be obtained in semiconductors with low effective electron mass $m^*\ll m$, for which the cyclotron frequency scales as $\wc\propto1/m^*$ \cite{ZF16,LB17}. In the presence of a planar surface parallel to the magnetic field, the torque is increased and a lateral force emerges due to AM conservation, even in the absence of rotation, an effect that could be observed through the lateral deflection of neutral particles incident on a planar surface exposed to an in-plane magnetic field. We also note that cosmic dust could be a potentially suitable testbed for these ideas, as it contains submicron particles exposed to a large range of vacuum temperatures and magnetic fields for very long periods of time. For example, gigantic magnetic fields are generated near stars, and in particular neutron stars. In this respect, the resulting nonlinear Zeeman effect could reveal additional physics in connection with vacuum friction.

\acknowledgments

This work has been supported in part by the Spanish MINECO (MAT2017-88492-R and SEV2015-0522), ERC (Advanced Grant 789104-eNANO), the Catalan CERCA Program, and Fundaci\'o Privada Cellex.

\appendix

\section{Transformation of the particle Hamiltonian to the rotating frame}
\label{A1}

For the sake of concreteness, we consider the electronic states of a particle described by the Hamiltonian $\mathcal{H}_{\rm part}(\{\rb_i\})=-\sum_i\hbar^2\nabla_i^2/2\me+e^2\sum_{i>i'}1/|\rb_i-\rb_{i'}|-e\sum_i V(\rb_i)$ when it is not rotating, where $\rb_i$ and $\rb_{i'}$ run over electron coordinates, while $V(\rb)$ is the potential produced by the atomic nuclei. The resulting many-body eigenstates $\psi_l(\{\rb_i\},t)$, labeled by the index $l$, satisfy the Schr\"odinger equation $\mathcal{H}_{\rm part}\psi_l=\ii\hbar\partial_t\psi_l$, and their time dependence is fully captured by $\psi_l(\{\rb_i\},t)=\psi_l(\{\rb_i\})\exp(-\ii\varepsilon_l t)$, where $\hbar\varepsilon_l$ is the state energy. When the particle is rotating, the atomic potential acquires a time dependence in the lab frame, which is trivially eliminated in the rotating frame (i.e., the nuclei appear to be frozen in the rotating frame), defined by the transformation $\rb=(r,\varphi,z,t)\rightarrow\rb'=(r',\varphi',z',t')=(r,\varphi-\Omega t,z,t)$, so that the lab-frame Hamiltonian remains the same if we substitute $\rb_i$ by $\rb'_i$ in it. However, in the right-hand side of the Schr\"odinger equation we have to substitute $\partial_t\rightarrow\partial_{t'}-\Omega\sum_i\partial_{\varphi'_i}$, and therefore, the rotating-frame Hamiltonian becomes  $\mathcal{H}_{\rm part}(\{\rb'_i\})-\Omega\mathcal{L}_z$, where $\mathcal{L}_z=-\ii\hbar\sum_i\partial_{\varphi'_i}$ is the many-body AM operator. Now, we focus on particles that are isotropic in the plane perpendicular to the rotation axis $z$ (i.e., with axial symmetry around that axis), and consequently, the above stationary eigenstates of the nonrotating particle can also be used to construct the stationary states of the rotating particle in the rotating frame as
\[\psi_l(\{\rb'_i\},t)=\psi_l(\{\rb'_i\})\ee^{-\ii(\varepsilon_l-m_l\Omega) t},\]
where we choose a basis set of eigenstates of $\mathcal{L}_z$ with eigenvalues $\hbar m_l$. Importantly, this is the appropriate choice of eigenstates that also diagonalizes the particle Hamiltonian in the presence of a magnetic field along $z$.

\section{Polarizablity of a rotating particle}
\label{secpola}

Linear response theory \cite{PN1966} provides us with an expression for the atomic polarizability in terms of particle eigenstate energies and transition dipoles. Because dipole components along $z$ are unaffected by particle rotation, it is clear that the polarizability remains unchanged for polarization along that direction, and additionally, there are not off-diagonal terms that mix $z$ with $x$ or $y$. We focus next on the polarizability tensor in the remaining $x$-$y$ subspace, which we obtain following the standard procedure of perturbing an initial state of the particle $|l\rangle$ under the influence of an external electric field, constructing the induced dipole as a result of this perturbation, and then averaging over initial state populations $f_l$ \cite{PN1966}. When following this procedure, we must take into account the rotation of the particle by transforming the external field to the rotating frame, where the states are defined as described in Appendix\ \ref{A1}, and then transforming back the resulting induced dipole to the lab frame. After some lengthy but straightforward algebra, we obtain
\begin{align}
\bar{\bar\alpha}(\omega)=\frac{1}{\hbar}\sum_{ll'} (f_{l'}-f_l)\,\frac{\pb^*_{l'l}\otimes\pb_{l'l}}{\omega-\varepsilon_{l'l}+\ii 0^+},
\nonumber
\end{align}
which is the general result for the polarizability of a particle, but in which we find that the populations $f_l\propto\ee^{-\hbar(\varepsilon_l-m_l\Omega)/\kB T_1}$ for the particle at temperature $T_1$ are determined by the frequencies $\varepsilon_l-m_l\Omega$ in the rotating frame, while the frequency differences $\varepsilon_{l'l}=\varepsilon_{l'}-\varepsilon_l$ are those of the particle at rest. Additionally, the transition dipole moments $\pb_{l'l}=-e\langle l'|\rb|l\rangle$ are also those of the particle at rest.

Considering that we are choosing $|l\rangle$ to be eigenstates of the AM operator $\mathcal{L}_z$ with eigenvalues $\hbar m_l$, it is convenient to express the dipole moments in the Zeeman coordinate basis defined in Sec.\ \ref{interaction}, which permits us to write $\pb^*_{l'l}\otimes\pb_{l'l}=(p^\perp_{l'l})^2\,\left(\uu_+\otimes\uu_-\delta_{m_{l'},m_l+1}+\uu_-\otimes\uu_+\delta_{m_{l'},m_l-1}\right)+\zz\otimes\zz (p^z_{l'l})^2$. Using these expressions, the polarizability tensor in the Zeeman basis reduces to
\[ \bar{\bar\alpha}(\omega)=\alpha_+(\omega)\,\uu_+\otimes\uu_- +\alpha_-(\omega)\uu_-\otimes\uu_+ + \alpha_0(\omega) \zz\otimes\zz,\]
where
\begin{align}
\alpha_\pm(\omega)&=\frac{1}{\hbar}\sum_{ll'} (f_{l'}-f_l) \delta_{m_{l'},m_l\pm1}\frac{(p^\perp_{l'l})^2}{\omega-\varepsilon_{l'l}+\ii 0^+}, \nonumber\\
\alpha_0(\omega)&=\frac{1}{\hbar}\sum_{ll'} (f_{l'}-f_l) \delta_{m_{l'},m_l}\frac{(p^z_{l'l})^2}{\omega-\varepsilon_{l'l}+\ii 0^+}. \nonumber
\end{align}
Interestingly, from the assumption of a Boltzmann distribution for $f_l$ (see above), we find the result
\begin{align}
&{\rm Im}\{\alpha_\pm(\omega)\}= \label{Imalpha}\\
&\frac{\pi/\hbar}{n_1(\omega\mp\Omega)+1} \sum_{ll'} f_{l}\,(p^\perp_{l'l})^2 \delta_{m_{l'},m_l\pm1}\,\delta\left(\omega-\varepsilon_{l'l}\right) \nonumber,
\end{align}
where $n_1(\omega)=1/(\ee^{\hbar\omega/k_\text{B}T_1}-1)$ is the Bose-Einstein distribution at the particle temperature $T_1$. We use this identity in the derivation of Eq.\ (\ref{AB}) in the main text. Incidentally, the identity $f_{l'}/f_l=\exp\left[\hbar(\omega\mp\Omega)/k_\text{B}T_1\right]$ holds inside the sum of Eq.\ (\ref{Imalpha}), which is useful in the derivation of Eqs.\ (\ref{AB}) and (\ref{EM}).


\end{document}